\documentclass[12pt]{article}
  \usepackage{graphicx}
\begin{document}
\input epsf
\def\be{\begin{equation}}
\def\bea{\begin{eqnarray}}
\def\ee{\end{equation}}
\def\eea{\end{eqnarray}}
\def\d{\partial}
\def\la{\lambda}
\def\eps{\epsilon}
\newcommand{\dm}{\begin{displaymath}}
\newcommand{\edm}{\end{displaymath}}
\renewcommand{\b}{\tilde{B}}
\newcommand{\gm}{\Gamma}
\newcommand{\ac}[2]{\ensuremath{\{ #1, #2 \}}}
\renewcommand{\ell}{l}
\def\bb{$\bullet$}

\def\q{\quad}

\def\bn{B_\circ}

\let\a=\alpha \let\b=\beta \let\g=\gamma \let\d=\delta \let\e=\epsilon
\let\z=\zeta \let\c=\chi \let\th=\theta  \let\k=\kappa
\let\l=\lambda \let\m=\mu \let\n=\nu \let\x=\xi \let\r=\rho
\let\s=\sigma \let\t=\tau
\let\vp=\varphi \let\vep=\varepsilon
\let\w=\omega      \let\G=\Gamma \let\D=\Delta \let\Th=\Theta
              \let\P=\Pi 

\renewcommand{\theequation}{\arabic{section}.\arabic{equation}}
\newcommand{\newsection}[1]{\section{#1} \setcounter{equation}{0}}

\def\h{{1\over 2}}
\def\t{\tilde}
\def\r{\rightarrow}
\def\nn{\nonumber}
\let\bm=\bibitem

\let\pa=\partial

\vspace{20mm}
\begin{center}
{\LARGE  Perturbations of supertube in KK monopole background}
\\
\vspace{20mm}
{\bf Yogesh K. Srivastava \footnote{yogesh@pacific.mps.ohio-state.edu}
\\}
Department of Physics,\\ The Ohio State University,\\ Columbus,
OH 43210, USA\\
\vspace{4mm}
\end{center}
\vspace{10mm}
\begin{abstract}

We study perturbations of supertube in KK monopole background, at both DBI and supergravity levels. We analyse both NS1-P as well as
D0-F1 duality frames and study different profiles. This illuminates certain aspects of bound states of KK monopoles with supertubes.
\end{abstract}
\newpage

\section{ Introduction}

Supertubes are $1/4$ supersymmetric bound states with $D0$ and $NS1$ brane charges as true charges, along with $D2$-brane as dipole charge. In flat space, Mateos and Townsend \cite{Mateos}
first constructed
supertubes by using Dirac-Born-Infeld (DBI)\footnote{We will refer various worldvolume actions as DBI even though for strings it would be Polyakov or 
Nambu-Goto action.}effective action for $D2$ brane and turning on worldvolume electric and magnetic fields.  
Branes corresponding to net charges, $NS1$ and $D0$, are represented as electric and magnetic fluxes on  $D2$ brane worldvolume or equivalently, these fields are due to `dissolved' $NS1$ and $D0$ branes in $D2$ 
brane worldvolume. $D2$-brane itself carries no net charge but only a dipole charge. Crossed electric and magnetic fields generate Poynting angular momentum which prevents the $D2$ brane from collapsing due to it's 
tension. Another way to describe this is to say that $NS1$ and $D0$ branes expand to $1/4$ supersymmetric $D2$ branes by the addition of angular momentum.

By dualities\footnote{We refer to various systems by their true charges without refering to dipole charges explicitly.}, this $D0$-$NS1$ system can be related to $NS1$-$P$ system, which is given by a $NS1$ string wrapped $n_1$ times around a circle $S^1$ and carrying  $n_p$ 
units of momentum along $S^1$ direction. Initial supertubes were constructed with  circular cross-section  but it was soon realized \cite{Mateos2,Bak}  that supertubes exist for any arbitrary
profile. This fact is a bit obscure in $D0$-$NS1$ language but in the $NS1$-$P$ duality frame, it's just a string carrying a right moving wave with
 an arbitrary profile. In \cite{Giusto} we studied adding perturbations to BPS supertubes (both in $D0$-$NS1$ and $NS1$-$P$ duality frames) and found classical 
solutions at both linear and non-linear levels (in the $NS1$-$P$ language these are just vibrations of fundamental string and it's trivial to write down the 
full solution). Based on several evidences, we formulated a conjecture which allows us to distinguish bound states from unbound states. The conjecture says that
 bound states are characterized by the absence of `drift' modes where by `drift' modes, we mean slow motion on moduli space of configurations. So
 when we have motion on moduli space we take the limit of the velocity going to zero, and  over a 
long time $\Delta t$ the system configuration changes by order unity. Using $\Delta x$ as a general symbol for the change in the configuration we have for `drift' on moduli space'
\be
v\sim \epsilon, ~~~\Delta t\sim {1\over \epsilon}, ~~~\Delta x\sim 1, ~~~~~~~(\epsilon\r 0)
\label{driftq}
\ee
On the other hand for the periodic behavior of bound states, we have
\be
v\sim \epsilon, ~~~\Delta t\sim 1, ~~~\Delta x\sim \epsilon ~~~~~~~(\epsilon\r 0)
\label{quasiq}
\ee

In the DBI description of supertubes, backreaction of branes on spacetime is neglected. Supergravity solution for two-charge supertube was constructed in
\cite{sugrasut} and in different duality frames in \cite{bal,mm,lm3}. It turns out that these correspond to simplest of microstate solutions for $2$-charge systems
 in $5$-dimensions and is completely smooth and horizon-free in $D1$-$D5$ duality frame. Based on AdS/CFT correspondence for $D1$-$D5$ system and several
other evidences, the Mathur conjecture \cite{lm4} is a proposal to associate \emph{bound} states in CFT to smooth,horizon-free geometries (whenever supergravity 
description is possible). The idea is 
to associate coherent state in the CFT with an asymptotically flat geometry which is smooth, free of horizons, 
carries the same conserved charges as the black hole, and hence constitutes a microstate of the black hole. 
 For the case of $2$-charge systems, all bosonic solutions were constructed by Mathur and Lunin \cite{lm4,Lunin}. 
Geometries with both bosonic and fermionic condensates were considered in \cite{taylor1} and relationship between gravity and CFT sides has been further 
explored in \cite{taylor2,taylor3} recently. Our understanding of the $3$-charge systems is less complete but a few examples are known \cite{Lunin2,gms1,gms2,bena2,levi}. In four dimensions, smooth solutions for $3$ and $4$-charges have appeared in 
the literature \cite{bena1,ash}.

Mathur conjecture emphasizes that microstates of black holes (which are described by smooth, horizon-free geometries when classical supergravity description is
possible) correspond to \emph{bound} states only. Hence it becomes very important to have a criteria to distinguish bound states from unbound ones as developed in
\cite{Giusto}. Recently, there has been much interest in studying solutions containing KK monopole \cite{bena1,ash,bena2}. Also, as recent work shows, it can be used to connect black rings in five dimensions
to black holes in four dimensions \cite{gaiot,elvang1}. Studies of black rings \cite{elvang2}  in Taub-NUT space \cite{bena1,bena2} led to supersymmetric solutions carrying angular momentum in four dimensional asymptotically flat space \cite{elvang1}. 
KK monopoles also occur in $4$-dimensional string theoretic systems \cite{self,bena1,ash,Bala} . Simplest of these configurations \cite{bena1,elvang1} can be considered as supertubes in KK monopole
background.

Unlike the case of supertubes\footnote{In $D0$-$NS1$ language, both charges were induced in a single higher brane $D2$ or 
in the $NS1$-$P$ language these are just vibrations of fundamental string.}where system was obviously bound, in the systems with KK monopole it is not
 a priori obvious that we are considering \emph{bound} states. Since we have a conjectured test which can distinguish, at least in principle, bound states from unbound ones, we would like to apply it 
to some of the systems with KK monopole. This involves solving perturbation equations in KK monopole background. Due to non-trivial background, 
non-linear perturbation equations, even at DBI level, are quite difficult to solve and hence we would restrict ourselves to linearized perturbations.
Thus our considerations are geared to analyze some of these systems, especially one corresponding to geometry
given in \cite{bena1} and to study it's boundedness properties using conjecture of \cite{Giusto}. We found that there are no `drift' modes at the DBI level
and system shows `quasi-oscillations' as discussed in \cite{Giusto}.  On the gravity side, we construct near ring limit
of the geometry and we were able to show that near ring limit is identical to near ring limit of $2$-charge systems considered \cite{Giusto} except for 
the periodicity of the ring circle. Then we consider torus perturbations as in \cite{Giusto} and find that results agree with DBI analysis.

\subsection{Outline of the paper}

Plan for the present paper is as follows.
\begin{itemize}
\item In $\S2$, we set up equations which describe motion of classical string in a general curved background and then apply them for the KK monopole background. We linearize about BPS solution of a
 fundamental string carrying a right moving wave and write down perturbation equations. 
\item In  $\S3$, we write the string profile corresponding to metric of Bena and Kraus (BK)  \cite{bena1}. In our system of $3$-charges in four 
dimensions different geometries correspond to different profile functions \cite{ashunpubl} of a one dimensional string, as was also the case for $2$-charge systems. Here we will work in $NS1-P$ duality frame. Then we solve perturbation equations  for this profile.
\item In  $\S4$, we discuss the construction of $D0$-$NS1$ supertube in KK monopole background. Then we study perturbations to this.
\item In $\S5$, we study a profile different from BK profile in both  $NS1$-$P$ language and $D0$-$NS1$ language.

\item In $\S6$, we study the supergravity side of the system. We take near ring limit of BK geometry and show how the perturbation analysis can be reduced to one done 
in a previous paper \cite{Giusto} and consequently time-period of torus vibrations also matches with the one considered in \cite{Giusto}. 

\item In $\S7$, we conclude with a discussion of our results and directions for future investigations.

\end{itemize}

\section{Oscillating string in KK monopole background}

In this section, we consider Polyakov string in KK monopole background. We first set up  equations of motion and 
 constraint equations in a general background. Our action is
\be
S= -\frac{T_1}{2} \int d\sigma d\tau \sqrt{g}g_{\alpha\beta}(\sigma,\tau)G_{AB}(X)\partial^{\alpha}X^{A}\partial^{\beta}X^B
\ee

Here $\sigma,\tau$ are worldsheet coordinates and $\alpha,\beta$ are worldsheet indices. Index $A$ for spacetime 
coordinates $X^A$ goes from $0,..9$. For worldsheet metric $g_{\alpha\beta}$ we have $g= -det(g_{\alpha\beta})$. 
Varying the action with respect to coordinates $X^A$, we get 
\be
\frac{\delta S}{\delta X^A}=0 = \partial_\alpha [\sqrt{g}G_{AB}\partial^\alpha X^B]- \frac{1}{2}(\partial_A G_{CD})\partial
_\alpha X^C \partial ^\alpha X^D
\ee 

In the conformal gauge on worldsheet, we have $g_{\alpha\beta}= e^{2f}\eta_{\alpha\beta}$. So we get
\be
\partial_\alpha[(\partial^\alpha X^B )G_{AB}]- \frac{1}{2}(\partial_A G_{CD})(\partial_{\alpha}X^C \partial^\alpha X^D)=0
\ee

Contracting with $G^{AP}$, we get
\be
\partial_\alpha \partial^\alpha X^P + G^{AP}[(\partial_C G_{AB})\partial^\alpha X^C \partial_\alpha X^B - \frac{1}{2}(\partial_A
G_{CD} \partial^\alpha X^C \partial_\alpha X^D] =0
\ee

In \cite{vega} (see also the references given there), general string equations of motion in curved background were given in a slightly different form. 
To match with those, we write the combination of derivatives as christoffel symbols. Writing 
\begin{displaymath}
(\partial_C G_{AB})\partial^\alpha X^C \partial_\alpha X^B = \frac{1}{2}[\partial_C G_{AB}+ \partial_B G_{AC}]\partial^\alpha
X^C \partial_\alpha X^B
\end{displaymath}

and recognising the combination of derivatives as christoffel symbols, we get
\be
\partial_\alpha \partial^\alpha X^P + \Gamma^{P}_{CB}\partial^\alpha X^C \partial_\alpha X^B = 0
\ee

Constraint equations are given by 
\be
\frac{\delta S}{\delta g^{\alpha\beta}}=T_{\alpha\beta}= G_{AB}[\partial_\alpha X^A \partial_\beta X^B - \frac{1}{2}g_{\alpha
\beta}\partial_\gamma X^A \partial^\gamma X^B]=0
\ee

If we choose lightcone variables $\xi^\pm=\t\tau\pm\t\sigma$ then $g_{++}=g_{--}=0$ and we get
\begin{eqnarray}
\partial_+ \partial_- X^A + \Gamma^{A}_{BC}\partial_+X^B \partial_-X^C = 0  \label{eom} \\
G_{AB}\partial_{\pm}X^A \partial_{\pm}X^B =0 \label{const}
\end{eqnarray}

$\pa_{\pm}$ denotes derivative with respect to $\xi^\pm $ in previous equation.

Ten dimensional metric for KK monopole at origin is 
\be
ds^{2} = -dt^{2} +dy^{2} + \sum_{i=6}^{9}dz^{i}dz_{i} + V[ ds + \chi_{j}dx^{j}]^{2} + V^{-1}[dr^{2} + r^{2}(d\theta^{2} +
\sin^{2}\theta d\phi^{2})] \label{kkmetric}
\ee
\be
V^{-1}= 1+ \frac{Q}{r} \ \ , \ \ \vec{\nabla} \times \vec{\chi} = - \vec{\nabla}V^{-1}
\ee
Here $y$ is compact with radius $R_{5}$ while
$x_{j}$ with $j=1,2,3$ are transverse coordinates while $z_{i}$ with $i=6,7,8,9$ are coordinates for torus $T^{4}$. Here 
$Q = \frac{1}{2} N_K R_K$ where $N_K$ corresponds to number of KK monopoles. Near $r=0$, $s$ circle shrinks to zero. For $N_K =1$, it does so smoothly
while $N_K >1$, there are $Z_{N_K}$ singularities. Here we just consider $N_K =1$ case. General problem of classical string propagation in KK 
monopole background is quite difficult to solve and so we will restrict ourselves to 
considering linearized perturbations about a given string configuration satisfying equations of motion. Our base configuration (about which we want to 
perturb) is fundamental string wrapped along $y$-circle and carrying a right moving wave or in other words, supertube in $NS1$-$P$ duality frame. 
 We know that in this case the waveform travels with the speed of light in the $y$ direction. Let us check that this is a solution of our string equations. This time we know the solution in the {\it static} gauge on the worldsheet: 
\be
t= b \t\tau, ~~~y=b\t\sigma
\label{static}
\ee
Writing $\t\xi^\pm=\t\tau\pm\t\sigma$ and noting that a right moving wave is a function of $\t\xi^-$ we expect the following to be a solution

\be
t= b\frac{\tilde{\xi}^+ + \t\xi^-}{2} \ \ , \ \ y= b\frac{\tilde{\xi}^+ -\t\xi^-}{2} \ \ , \ \ X^{\mu}= x^{\mu}(\t\xi^-)
\ee

We see that this satisfies equation of motion. But it doesn't satisfy constraint equations i.e the induced metric on worldsheet is not conformal to flat
metric. 
\be
ds^2 = - b^{2} d\tilde{\xi}^+ d\t\xi^{-} + G_{\mu\nu}(x'^{\mu}x'^{\nu})(d\t\xi^{-})^{2}
\ee

where primes denote differentiation wrt. $\t\xi^{-}$. However, as done in $\cite{Giusto}$, we change coordinates to 
\be
(\xi^{+},\xi^-)= (\tilde{\xi}^+ - f(\t\xi^-),\t\xi^-) \label{change1}
\ee

with 
\be
f'(\xi^-)= \frac{G_{\mu\nu} x'^{\mu}x'^{\nu}}{b^2} \label{change2}
\ee

Here prime now denotes derivative with respect to $\xi^-$ and index $\mu$ denotes directions along taub-nut part of KK monopole.
 In terms of these new coordinates, we have a conformally flat metric on the worldsheet. 
\be
ds^2=-b^2\,d\xi^-\,d\xi^+
\ee

So configuration
\be
t= b\frac{\xi^+ + \xi^- + f(\xi^-)}{2} \ \ , \ \ y= b\frac{\xi^+ -\xi^- + f(\xi^-)}{2} \ \ , \ \ x^{\mu}= x^{\mu}(\xi^-)
\ee

satisfies the equations of motion. For $t,y$ coordinates, there is separation between left and right movers and hence equation of motion is trivially satisfied.
We will take vibrations along torus coordinates to be zero in the base configuration. 
For coordinates along taub-nut, christoffel symbols are non-zero but since $\partial_{+}x^{\mu}(\xi^{-})=0$, equations of 
motion are satisfied. This is to be expected since KK monopole is an exact background for string theory. Now we consider 
linearized perturbations about this configuration 
\be
X^{\mu}= x^{\mu}(\xi^-) + \epsilon Y^{\mu}(\xi^+ , \xi^-)  \  \  , \  \  z_j = \epsilon Z_j (\xi^+,\xi^-)
\ee

where $\epsilon$ is a small parameter. We will neglect terms of higher order in $\epsilon$ in what follows. 
In a curved spacetime, left and right movers are mixed and hence $Y^\mu$ depends on both $\xi^\pm$. Expanding eqn. \ref{eom} to first order in $\epsilon$ perturbation 
equation will be
\be
\partial_+ \partial_- Y^\mu + \Gamma^{\mu}_{\nu\rho}\partial_+Y^\nu \partial_-x^\rho = 0 
\ee

Here christoffel symbols are calculated using zeroth order background metric evaluated for the base configuration and hence it depends only on $\xi^-$. 
Hence we have following first order equation after first integration.
\be
\partial_- Y^{\mu} + \Gamma^{\mu}_{\nu\rho}Y^\nu \partial_-x^\rho = h^{\mu}(\xi^-) 
\label{feom}
\ee

For directions $z_j$, christoffel symbols are zero and hence perturbations  are of the form 
\be
Z_j = Z_{-}(\xi^-) + Z_{+}(\xi^+)
\ee

Before trying to solve the equation \ref{feom} , let us try to see the form of constraint equations \ref{const} for these solutions. First Constraint
equation becomes
\be
0=G_{AB}\partial_+ X^A \partial_+ X^B = -\frac{b^2}{4} + \frac{b^2}{4} + \epsilon^2 G_{\mu\nu}\partial_+ Y^{\mu}\partial_+ Y^{\nu}
+  \epsilon^2 G_{ji}\partial_+ Z^{j}\partial_+ Z^{i}
\ee

and second non-trivial one becomes
\begin{eqnarray}
G_{AB}\partial_- X^A \partial_- X^B = -\frac{b^2}{4}(1+ f'(\xi^-))^2 + \frac{b^2}{4}(-1+f'(\xi^-))^2 +  \epsilon^2 G_{ji}\partial_+ Z^{j}\partial_+ Z^{i} \nonumber \\
 + G_{\mu\nu}\partial_- X^{\mu}\partial_- X^{\nu} =0
\end{eqnarray}                        

We see that upto first order in $\epsilon$, first constraint is satisfied. Now we manipulate order $\epsilon$ terms in second equation a bit to 
get the constraint on $h^{\mu}$ implied by the second equation. We first expand all the terms into base quantities and perturbations.
\be
G_{\mu\nu}\partial_- X^{\mu}\partial_- X^{\nu}= (\overline{G}_{\mu\nu} + \epsilon h_{\mu\nu})(x'^{\mu}+ \epsilon Y'^{\mu})
(x'^{\nu}+ \epsilon Y'^{\nu})
\ee

Here $\overline{G}_{\mu\nu}$ is the four dimensional taub-nut part of base metric evaluated for base configuration $x^{\mu}$ and $h_{\mu\nu}$ is the 
linearized perturbation in metric. Prime denotes derivative with respect to $\xi^-$. Putting this in constraint equation and considering terms upto order $\epsilon$ only, we get
\be
-b^2f'(\xi^-)+ \overline{G}_{\mu\nu}x'^{\mu}x'^{\nu}+ \epsilon [h_{\mu\nu}x'^{\mu}x'^{\nu} + 2\overline{G}_{\mu\nu}x'^{\mu}Y'^{\nu}]=0
\ee

At zeroth order in $\epsilon$, terms vanish by the definition of $f(\xi^-)$. To further massage first order terms, we put $Y'^\mu$ from the equation of motion in 
the constraint equation. 
\be
[h_{\mu\nu}x'^{\mu}x'^{\nu} + 2\overline{G}_{\mu\nu}x'^{\mu}(-\Gamma^{\nu}_{\rho\sigma}Y^\rho x'^{\sigma} + h^\nu (\xi^-) ]=0 \label{const1}
\ee

Here we have set 
\be
-b^2f'(\xi^-)+ \overline{G}_{\mu\nu}x'^{\mu}x'^{\nu}=0 \label{fdef}
\ee

giving $f'(\xi^-)$. Putting the definition of christoffel symbols , we get 
\be
2\overline{G}_{\mu\nu}x'^{\mu}(\Gamma^{\nu}_{\rho\sigma}Y^\rho x'^{\sigma})=  x'^{\mu}x'^{\sigma}Y^{\rho}\overline{G}_{\mu\nu}
\overline{G}^{\nu \alpha}(\partial_\rho \overline{G}_{\alpha\sigma}+ \partial_\sigma \overline{G}_{\alpha\rho}-
\partial_\alpha \overline{G}_{\rho\sigma})= x'^{\mu}x'^{\sigma}Y^{\rho}\partial_\rho \overline{G}_{\mu\sigma}
\ee

Putting this in eqn. \ref{const1} we get
\be
x'^{\mu}x'^{\sigma}(h_{\mu\sigma}- Y^\rho \partial_\rho \overline{G}_{\mu\sigma}) + 2\overline{G}_{\mu\nu}h^\nu x'^{\mu} =0
\ee

This gives constraint on $h^\nu$ as first term automatically vanishes. Thus
\be
\overline{G}_{\mu\nu}h^\nu x'^{\mu} =0 \label{constraint}
\ee

is the final form of constraint equation which we will use later. 
\section{Perturbations for BK  string profile}

Now we have set up our equations of motion and constraint equations. So we can use these to find linearized perturbations for given base configurations.
Our interest is in systems which correspond to supertubes in KK monopole background. Geometries constructed in \cite{bena1,bena2,elvang1} correspond to such situations.
In \cite{ashunpubl}, it is shown that geometry corresponding to Bena-Kraus (BK) metric is generated by considering a particular string profile in
KK monopole background and other string profiles give different geometries, generalizing those of \cite{lm4} ( where general geometries correspond to 
string profile in four dimensional flat space) to the case of $3$-charges in four dimensions. Since we finally want to consider supergravity 
perturbations in Bena-Kraus geometry, it would be necesary to consider same string profile (which generates BK geometry in supergravity limit) as our
base configuration about which we add perturbations. So in this section, we determine the profile corresponding to BK geometry in coordinates 
appropriate for KK monopole background. Since taub-nut space becomes flat space for small distances, we can find the profile corresponding to BK metric
 by considering the profile near the center of KK monopole. For this we need conversion between flat space coordinates and taub-nut coordinates.
Since we have $\cos\theta$ as the gauge field for taub-nut instead of usual $1-\cos\theta$, we give the calculation for 
our case. Taub-nut metric is 
\be
ds^2 = \frac{1}{V}(dz + \cos\theta d\phi)^2 + V(dr^2 + r^2 d\theta^2 + r^2 \sin^2 \theta d\phi^2)
\ee

Here $V= 1+ \frac{Q}{r} \label{V}$ and $0\leq \theta \leq \pi$. The periodicities of angular coordinates are $\delta\phi =2\pi$ and
$\delta z= 2\pi R_{K}$ where $Q= \frac{1}{2}N_K R_K$. Here $R_K$ is the asymptotic radius of $z$-circle and $N_K$ is the 
number of monopoles. For $r<<Q$, we have
\be
ds^2 \ \approx \ Q (\frac{dr^2}{r} +  r d\theta^2 + r\sin^2 \theta d\phi^2) + \frac{r}{Q}(dz+ A\cos\theta d\phi)^2
\ee

Now we make change of variables by defining 
\be
\rho = 2\sqrt{Qr} \ \ \ , \ \ \ \tilde{\theta}= \frac{\theta}{2}
\ee

Now metric becomes
\begin{eqnarray}
ds^2 = d\rho^2 + \rho^2 d\tilde{\theta}^{2} + \frac{\rho^2}{4}\sin^2 2\tilde{\theta} d\phi^2 + \frac{\rho^2}{4Q^2}(dz + Q
\cos 2\tilde{\theta} d\phi)^2  \\
 = d\rho^2 + \rho^2 d\tilde{\theta}^{2} + \frac{\rho^2}{4}d\phi^2 +\frac{\rho^2}{4}dz^2 + 
\frac{\rho^2}{2Q}\cos 2\tilde{\theta} d\phi dz
\end{eqnarray} 

Inserting $1= sin^2 \tilde{\theta} + \cos^2 \tilde{\theta}$, we get
\be
ds^2 = d\rho^2 + \rho^2 [ d\tilde{\theta}^2 + \frac{\cos^2 \tilde{\theta}}{4}(\frac{1}{Q}dz +\phi)^2 + \frac{\sin^2 \tilde{\theta}}{4}
(\frac{1}{Q}dz -\phi)^2 ]
\ee

We define following combinations
\be
2 \tilde{\psi}= \frac{1}{Q}dz +\phi \ \ \ , \ \ \ 2\tilde{\phi}= \frac{1}{Q}dz -\phi
\ee

In terms of these quantities, we have flat metric
\be
ds^2 = d\rho^2 + \rho^2[d\tilde{\theta}^2 + \cos^2\tilde{\theta}d\tilde{\psi}^2 + \sin^2\tilde{\theta}d\tilde{\phi}^2 ]
\ee

In this flat space metric, cartesian coordinates are defined by 
\be
x_1 = \rho \sin\tilde{\theta}\cos\tilde{\phi} , \ x_2 = \rho \sin\tilde{\theta}\sin\tilde{\phi}\ , \ 
x_3 = \rho \cos\tilde{\theta}\cos\tilde{\psi}\ , \  x_4 = \rho \cos\tilde{\theta}\sin \tilde{\psi} 
\ee  

BK metric for $3$-charges in $4$ dimensions is analogous to supertube metric \cite{sugrasut,Bala,mm,lm3} for $2$-charges in $5$ dimensions. Near the 
centre of KK monopole, we know that KK monopole metric reduces to flat space as we saw above. There string profile of BK metric must be same as
string profile of supertube metric. As show in \cite{lm3,Lunin}, we have circular profile function 
\be
F_1 = a \cos \omega v \ \ \ , \ \ \ F_2 = a \sin \omega v
\ee

for  geometry corresponding to simplest supertube. 
Here $v= t-y$.We see that we have following coordinates for profile function in polar coordinates
\be
\rho =a \ \ , \ \ \tilde{\theta}= \frac{\pi}{2} \ \ , \ \ \tilde{\phi}= \omega v 
\ee

The value of $\tilde{\psi}$ is indeterminate. To simplify things we take $\t\psi = \omega v$. In terms of taub-nut coordinates
 these values   translate to 
\be
r= \frac{a^2}{4Q} \ \ , \ \ \theta= \pi \ \ , \ \ \phi=0 \ \ , \ \ z=2Q\omega v  
\ee

Here $\omega= \frac{1}{nR_y}$ for the state we are considering, $n$ being the number of times string winds around $y$-circle of radius $R_y$.
\subsection{Perturbations of BK profile}

In section $2$, we determined the equation of motion \ref{feom} and constraint equation \ref{constraint} for a general base configuration
given by profile $x^{\mu}(\xi^-)$ in taub-nut directions. In this subsection, we apply these for the case of BK profile in $NS1$-$P$ duality frame.
From section $2$, equation of motion for perturbations $Y^A$ are 
\be
\partial_- Y^A + \overline{\Gamma}^{A}_{BC}Y^B \partial_-x^C = h^A (\xi^-)
\ee

In the flat directions $z_j$, solution is like in flat space i.e 
\be
Z_j = Z_{-}(\xi^-) + Z_{+}(\xi^+)
\ee

\be
t= b\frac{\xi^+ + \xi^- + f(\xi^-)}{2} \ \ , \ \ y= b\frac{\xi^+ -\xi^- + f(\xi^-)}{2} 
\ee

where  
\be
-b^2f'(\xi^-)+ \overline{G}_{\mu\nu}x'^{\mu}x'^{\nu} =0
\ee
gives $f'(\xi^-)$. In the taub-nut directions, we will only consider perturbations along $r$ and $z$. Since there are 
coordinate singularities at $\theta=\pi$, we can work at $\pi -\delta$ and then take \footnote{Since these coordinates have
singularities at $\theta=\pi$, we should change to other coordinate patch to cover the point $\theta=\pi$ (in that patch $\theta=0$ will have problem).
In other coordinate system, similar conclusions follow and hence we will not worry about these spurious singularities any further}
limit $\delta \rightarrow 0$. 
In what follows, we will set $z= R_K \psi$ to simplify some calculations. We will need following components of connection in what follows
\be
p= \overline{\Gamma}^{\psi}_{r\psi}= \frac{Q}{2r(Q+r)} \ \ , \ \ -q= \overline{\Gamma}^{r}_{\psi \psi}= -\frac{Q r R_{K}^{2}}
{2 (Q+r)^{3}}
\ee
We  consider the case where base configuration has non-constant radius. We consider taub-nut directions as 
perturbations in other directions ( whose connection components vanish ) are same as above. Base configuration
 in this case is 
\be
r= R(\xi^-)  \ \ , \ \ \theta= \pi \ \ , \ \ \phi=0 \ \ , \ \  \psi = \frac{2\omega Q}{R_K} \xi^- = \alpha \xi^-
\ee

We consider perturbations only along $r,\psi$ directions. Then 
\be
\partial_-\overline{X}^{\psi}= \alpha \ \ \ \ ,  \ \ \ \  \partial_-\overline{X}^{r} = R'= \frac{dR(\xi^-)}{d\xi^-}
\ee

Putting these in 
\be
\partial_- Y^A + \overline{\Gamma}^{A}_{BC}Y^B \partial_-x^C = h^A (\xi^-)
\ee

we get following two equations for the perturbations
\bea
\partial_-Y^r + \overline{\Gamma}^{r}_{rr} Y^r R' + \overline{\Gamma}^{r}_{\psi\psi}\alpha Y^\psi = h^{r}(\xi^-) \\
\partial_-Y^\psi + \overline{\Gamma}^{\psi}_{r\psi} \alpha Y^r + \overline{\Gamma}^{r}_{\psi r}R' Y^\psi = h^{\psi}(\xi^-)
\eea

Apart from equation of motion, we also have constraint equation \ref{constraint} which give following relation between $h^r$ and $h^{\psi}$.
\be
\overline{G}_{rr} h^{r} R' + \overline{G}_{\psi\psi}h^{\psi}\alpha =0
\ee

Putting the values of appropriate connection components, we get
\bea
\partial_-Y^r -\frac{QR'}{2R(Q+r)} Y^r  -\frac{QR\alpha}{2(Q+R)^3} Y^\psi = h^r(\xi^-) \\
\partial_-Y^\psi + \frac{Q}{2R(Q+R)} (\alpha Y^r + R' Y^\psi)  = h^\psi(\xi^-) 
\eea

 Multiplying the first equation by $\sqrt{V}$, dividing the second by $\sqrt{V}$ and using expression \ref{V} for $V$, we can write the two equations as
\bea
\partial_-( \sqrt{V}Y^r)  -\frac{Q\alpha}{2(Q+R)^2}\left( \frac{Y^\psi}{\sqrt{V}}\right)  = (\sqrt{V}h^r)(\xi^-) \\
\partial_- \left(\frac{Y^\psi}{\sqrt{V}}\right) + \frac{Q\alpha }{2(Q+R)^2}(\sqrt{V}Y^r)  = \left(\frac{h^\psi}{\sqrt{V}}\right)(\xi^-) 
\eea

Defining new dependent and independent variables
\bea
\tilde{Y}^r = \sqrt{V}Y^r \ \ \ , \ \ \ \tilde{Y}^{\psi}= \frac{Y^\psi}{\sqrt{V}} \\
\tilde{\xi}= \frac{Q\alpha}{2} \int \frac{d\xi^-}{(Q+R(\xi^-))^2}
\eea 

we get
\bea
\t\partial_-(\tilde{Y}^r)  - \tilde{Y}^\psi  = G^r (\t \xi) \\
\t\partial_-(\tilde{Y}^\psi)  + \tilde{Y}^r  = G^\psi (\t \xi)
\eea

where $\t\partial_-$ denotes derivative with respect to $\t\xi$ and new arbitrary functions $G^{r},G^{\psi}$ are now expressed as functions of $\t\xi$.
Since we will be needing it later also, let us solve equations of motion in a general form. We can express the above coupled first order inhomogeneous equations as matrix equation
\be
 \t\partial \vec{Y} = {\bf{A}}\vec{Y} + \vec{G} \label{meqn}
\ee

where ${\bf{A}}$ is a $2\times2$ matrix
\be
{\bf{A}} = \left(\begin{array}{clcr}
0 & 1 \\
-1 &  0
\end{array} \right)
\ee 

Solution to matrix equations like \ref{meqn} is found by diagonalizing the matrix ${\bf{A}}$. If $\lambda_j$ are eigenvalues and $\vec{S}_j$ are 
eigenvectors, with $j= 1,2$ then solution is given by 
\be
\vec{Y} = \sum_{j} c_j \vec{S}_j e^{\lambda_j \t\xi} + \sum_{j} e^{\lambda_j \t\xi}\vec{S}_j \int e^{-\lambda_j \t\xi}\t G_j(\t\xi)d\t\xi
\ee

Here $\vec{\t G} = S^{-1} \vec{G}$ and $S=[S_1 \ S_2] $ is the matrix of eigenvectors as column vectors. For our case, we get

\begin{eqnarray}
\left(\begin{array}{c} 
\t Y^r \\ \t Y^\psi
\end{array} \right)= c_{1}(\xi^+) \left(\begin{array}{c} 
i \\ 1
\end{array} \right)e^{i\t\xi} + c_{2}(\xi^+) \left(\begin{array}{c} 
-i \\ 1
\end{array} \right)e^{-i\t\xi}   \nonumber \\ 
+   e^{i\t\xi}\left(\begin{array}{c} 
i \\ 1
\end{array}\right)\int e^{-i\t\xi} \t G^r(\t \xi)d\t\xi + e^{-i\t\xi}\left(\begin{array}{c} 
-i \\ 1
\end{array}\right) \int e^{i\t\xi} \t G^\psi(\t\xi)d\t\xi
\eea

Solution can then be combined to schematically write down 
\bea
\tilde{Y}^r + i\tilde{Y}^\psi = B(\xi^+)e^{-i\tilde{\xi}} + G_1(\tilde{\xi}) \\
\tilde{Y}^r - i\tilde{Y}^\psi = A(\xi^+)e^{i\tilde{\xi}} + G_2(\tilde{\xi}) \label{solnfp}
\eea

\section{D0-F1 supertube in KKM background}

In previous section, we found perturbed solution corresponding to oscillating string in KK monopole background. We know that this 
system is dual to usual $D0-NS1$ supertube. In this section, we study supertube in $D0-NS1$ duality frame. Since we are in a non-trivial 
background (KK monopole), it is not clear how to do dualities required to go from $NS1$-$P$ to $D0$-$NS1$ frame as done in \cite{Giusto}. 
So we perform calculation of linearized perturbation separately in this duality frame. Static case of $D0$-$NS1$ supertube was 
considered in \cite{proeyen}. Here we will review their construction for the case of round supertube. In the next subsection we will add perturbations to it.

$D2$ supertube has world-volume coordinates $\sigma^0 ,\sigma^1 ,\sigma^2 =\sigma $. We embed supertube in such a way that  
\be
\sigma^0 =t \ \ , \ \ \sigma^1 =y \ \ , \ \  X^\mu = X^\mu (\sigma^2)
\ee

Here $X^\mu$ are arbitrary functions of $\sigma$. To stabilize the brane against contraction due to brane-tension, we introduce gauge field
\be
F = E d\sigma^0 \wedge d\sigma^1 + B(\sigma^2) d\sigma^1 \wedge d\sigma^2 = E dt\wedge dy + B(\sigma) dy \wedge d\sigma
\ee

For $D2$-brane of tension $T_2$, Lagrangian is given by
\be
\mathcal{L}= -T_2 \sqrt{-det[g+F]} = -T_2\sqrt{B^2 + R^2(1-E^2)}
\ee

Here $g$ is induced metric and $R^2 = G_{\mu\nu}X'^\mu X'^\nu $ and prime denotes differentiation wrt $\sigma$. Background metric $G_{\mu\nu}$ for 
KK monopole is given by \ref{kkmetric}. We define electric displacement as
\be
\Pi = \frac{\partial \mathcal{L}}{\partial E} = \frac{T_2 E R^2}{\sqrt{ B^2 + R^2(1-E^2)}}
\ee

In terms of this, we write hamiltonian density as
\be
\mathcal{H}= E\Pi -\mathcal{L} = \frac{1}{R}\sqrt{(R^2 + \Pi^2)(B^2 +R^2)}
\ee

It is easy to see that minimum value for $\mathcal{H}$ is obtained if $T_2 R^2 = \Pi B$ or $E^2 =1$. These conditions agree with what one gets from 
supersymmetry analysis. As in flat space, $B(\sigma)$ is an arbitrary function of $\sigma$. By the usual interpretation,
fluxes above correspond to $D2$ brane carrying both $D0$ and $F1$( along $y$ direction) charges. We are assuming isometry 
 along $y$-direction. Charges are given by
\begin{eqnarray}
Q_0 = \frac{T_2}{T_0}\int dyd\sigma B(\sigma) \\
Q_1 =  \frac{1}{T_1}\int d\sigma \Pi(\sigma) = \frac{T_2}{T_1}\int d\sigma\frac{E R^2}{\sqrt{ B^2 + R^2(1-E^2)}}
\end{eqnarray}

Round supertube in KK is given by
\be
\sigma^0 =t \ \ , \ \ \sigma^1 =y \ \ , \ \ R_0 = \frac{a^2}{4Q} \ \ , \ \ \theta= \pi \ \ , \ \ \phi=0 \ \ , \ \
z=2Q\omega\sigma  \label{profile} 
\ee

We have chosen parameters in such a way as to facilitate comparison with $NS1$-$P$ duality frame. In terms of flat space (or near the 
center of KK monopole), this corresponds to a circular profile in say, $(X_1,X_2)$ plane. We are not perturbing along torus directions. For this configuration 
\be
R^2 = X'^\mu X'_\mu = G_{zz}(2Q\omega)^2
\ee

So the supersymmetry condition gives a relationship between all three charges and the compactification radius. Now consider perturbation of this configuration.
\subsection{Perturbations in D0-F1 picture}

Let $R$ and $\sigma$ be the radial and angular coordinates in the $(X_1,X_2)$ plane. We choose the gauge $A_t =0$ for the worldvolume gauge field. Thus
 the gauge field has the form
\be
A = A_{\sigma}d\sigma + A_y dy \label{gaugefield}
\ee
\be
F = E dt\wedge dy + B dy\wedge d\sigma + \partial_t a_\sigma dt\wedge d\sigma
\ee
We want to study fluctuations around the configuration given by \ref{profile},\ref{gaugefield} with $\overline{E}=1$ and $b= \overline{B}(\sigma)$.
Lagrangian is given by
\be
\mathcal{L}= -T_2 \sqrt{-det[g+F]}
\ee

Putting values from \ref{profile},\ref{gaugefield}, we get
\be
 \mathcal{L}= -T_2\sqrt{(1-E^2)X'^2 -\dot{X}^2 X'^2 + (\dot{X}\cdot X')^2 - \dot{a}^2_\sigma + B^2 (1-\dot{X}^2) -2 EB \dot{X}\cdot X'} \label{lag}
\ee

We perturbed as
\begin{equation}
R= R_0 + \epsilon r(\sigma,t)  \ \ \ ,  \ \ \ E =1 + \epsilon \dot{a}_y \ \ \ , \ \ \ B = b -\epsilon a'_y
 \ \ \ , Z= \alpha\sigma + \epsilon z 
\end{equation}

where lower case quantities denote fluctuations. Field strength becomes
\be
F = (1 + \epsilon \dot{a}_y)  dt\wedge dy + (b -\epsilon a'_y) dy\wedge d\sigma + \partial_t a_\sigma dt\wedge d\sigma
\ee
We have put perturbations along $\theta,\phi$ directions to be zero. Putting these in Lagrangian and expanding upto second order

\be
\frac{\mathcal{L}}{T_2} = L^{(0)}+ \epsilon L^{(1)} + \epsilon^2 L^{(2)}
\ee

we find
\bea
L^{(0)} = -b \\
L^{(1)}= \frac{1}{b}\left(\frac{\alpha^2 \dot{a}_y}{V_0} + \frac{\alpha b \dot{z}}{V_0} + b\dot{a}_y \right)
\eea

We see that first order perturbation is a total derivative. This follows from the fact that our unperturbed configuration
satisfies equations of motion. Term second order in $\epsilon$ is given by
\dm
L^{(2)} = \frac{\alpha^2 + V_0 b^2}{2b} \dot{r}^2 + \frac{\alpha^2 (\alpha^2 + V_0 b^2)}{2b^3 V_0 ^2}\dot{a}_y^2 + \frac{(\alpha^2
+ V_0 b^2)}{2b V_0 ^2}\dot{z}^2 + V_0 \dot{r}r' + \frac{1}{V_0}\dot{z}z' + \frac{\alpha^2}{b^2 V_0}\dot{a}_y a'_y
\edm
\be
+ \frac{Q \alpha^2}{bV_0^2 R^2}r\dot{a}_y + \frac{\alpha(\alpha^2 + V_0 b^2)}{V_0^2 b^2}\dot{a}_y\dot{z} + \frac{\alpha Q}
{V_0^2 R^2}r\dot{z} + \frac{2\alpha}{b V_0}\dot{a}_y z'
\ee

From this , we get following equations of motion
\bea
\left(\frac{\alpha^2 + V_0 b^2}{b}\right)\ddot{r} + 2V_0\partial_\sigma \dot{r} - \frac{\alpha Q}{V_0^2 R^2}(\dot{z}+ \frac
{\alpha}{b}\dot{a}_y ) =0 \\
\left(\frac{\alpha(\alpha^2 + V_0 b^2)}{V_0^2 b^2}\right)[\ddot{z} + \frac{\alpha}{b}\ddot{a}_y ] + \frac{2\alpha}{V_0 b}
\partial_\sigma (\dot{z}+ \frac{\alpha}{b}a_y) + \frac{\alpha^2 Q}{b V_0^2 R^2}\dot{r}=0 \\
\left(\frac{(\alpha^2 + V_0 b^2)}{V_0^2 b}\right)[\ddot{z} + \frac{\alpha}{b}\ddot{a}_y ] + \frac{2}{V_0 }
\partial_\sigma (\dot{z}+ \frac{\alpha}{b}a_y) + \frac{\alpha Q}{V_0^2 R^2}\dot{r} =0
\eea

We see that second and third equations are same. If we define $x= z + \frac{\alpha}{b}a_y$ then we have following equations
\bea
\left(\frac{\alpha^2 + V_0 b^2}{b}\right)\ddot{r} + 2V_0\partial_\sigma \dot{r} - \frac{\alpha Q}{V_0^2 R^2}\dot{x} =0 \\
\left(\frac{\alpha^2 + V_0 b^2}{b}\right)\ddot{x} + 2V_0\partial_\sigma \dot{x} +\frac{\alpha Q}{ R^2}\dot{r} =0
\eea

We notice that as in the case of supertube in flat space, we only have time derivatives of field in the equations of motion. Thus any time 
independent perturbation is a solution, confirming that supertube in KK monopole background also has a family of time independent solutions. 
Solution to above equations is given by
\bea
r = c_1(\xi^+) \cos a\sigma + c_2(\xi^+)\sin a \sigma \\
x= k \left[ c_1(\xi^+) \sin a\sigma - c_2(\xi^+)\cos a \sigma \right] \label{dpert}
\eea

where
\be
\xi^+ = \frac{2t}{b} - \sigma -\frac{\alpha^2}{V_0 b^2}\sigma \ \ \ , \ \ \ k= -V_0  \ \ \ , \ \ \ a= \frac{\alpha Q}{V_0 ^2
R^2}
\ee

We see that here perturbations along $z$ direction add with the gauge field and only a combination occurs in equations
of motion. This occurs because as in $NS1$-$P$ duality frame, we have only two independent degrees of freedom. It's important to note that
 frequencies of oscillation agree in both the duality frames, as one expects. Motion is periodic and as in flat space case, we did not 
 find `drift' modes. So according to conjecture of \cite{Giusto}, this would correspond to bound state
 \subsection{Period of oscillation}

We had earlier defined  static gauge coordinates  $\t \tau$ and $\t\sigma$ in equation \ref{static} and then obtained conformal coordinates $\xi^+,\xi^-$
 from them using equations \ref{change1},\ref{change2} respectively. For finding the period of oscillation, it would be convenient to take 
 $\tilde{\tau}$ and $\xi^-$ as our basic variables. Relationship between target space time $t$ and $\t \tau$ is just by a simple multiplicative 
 factor $b$ while $\xi^-$ gives the parametrization of unperturbed base configuration. So in terms of these, we have
 \be
 \xi^+ = \t\xi^+ - f(\xi^-) = 2\t\tau -\xi^- -f(\xi^-)
 \ee
 
 The time dependence of the solution (\ref{dpert}) is contained in functions like $A(2\tilde\tau-\xi^--f(\xi^-))$ and similarly for torus directions
 $Z_j(2\tilde\tau-\xi^--f(\xi^-))$. We write 
\be
\xi^-+f(\xi^-)=\int_0^{\xi^-} d\chi (1+f'(\chi))
\ee
So the change in $\xi^-+f(\xi^-)$ when $\xi^-$ increases by $2\pi$ can be written as
$\int_0^{2\pi}d\chi (1+f'(\chi))$. We then find that the argument of $A, Z_j$ are unchanged when $(\t\tau, \xi^-)\r (\t\tau+\Delta \t\tau, \xi^- +2\pi)$ with
\be
2\Delta \tilde{\tau} -\int_{0}^{2\pi}
\left(1+ {\bar f}'(\bar{\xi}^-)\right)d\bar{\xi}^- =0
\ee

Using expression for $f'(\xi^-)$ as given in \ref{fdef}, we get
\be
\Delta \tilde{\tau} = \frac{1}{2}\int_{0}^{2\pi}\left(1+ \frac{\overline{G}_{\mu\nu}x'^{\mu}x'^{\nu}}{\bar {b}^{2}}
\right)d\bar{\xi}^- 
\ee

For torus directions, situation is similar to flat space case. In case of only torus vibrations, we get back flat space result
\be
\Delta t = \frac{1}{2}\left(\frac{M_{D0} + M_{NS1}}{M_{D2}}\right) \label{period}
\ee

\section{ Profile in 3-d part of KK}

Uptill now, we have considered BK profile only. In this section, we consider a different profile which seems natural for KK monopole background. 
We consider unperturbed profile with $z= \overline{X}_3 =0$ and perturbations only along $X_1 , X_2$ directions only, with $X_1 , X_2$ being arbitrary functions.
First consider the perturbations in $NS1$-$P$ language. Again we use equations of motion \ref{feom} and constraint equation \ref{constraint} as derived 
in section $2$ previously\footnote{We will denote directions along three dimensional part of taub-nut (i.e excluding fibre direction) by latin letters}.
\be
\partial_- Y^{i} + \Gamma^{i}_{jk}Y^\nu \partial_-x^k = h^{i}(\xi^-)
\ee

Relevant christoffel connections in this case are
\be
\Gamma^{i}_{jk} = \frac{1}{2V}\left[(\partial_j V)\delta^{i}_{k} + (\partial_k V)\delta^{i}_{j} -(\partial_l V)\delta^{il}
\delta_{jk} \right]
\ee

In other directions, christoffel symbols are zero and hence perturbation equations are trivial as shown in section $2$. Here we concentrate on 
fluctuations along three dimensional part of taub-nut which is conformal to flat space.
We put relevant connection coefficients in perturbation equation for taub-nut directions and get
\be
\partial_- Y^{i} + \frac{1}{2V}\left[(\partial_k V)\partial_-\overline{X}^{k}Y^{i} + (\partial_j V)Y^{j}\partial_-\overline{
X}^{i} -(\partial_l V)\delta^{il}\delta_{jk} \partial_- \overline{X}^k Y^j \right] = h^i(\xi^-)
\ee

Writing $S= ln V$ and 
\be
w^{i}= e^{\frac{1}{2}\int (\partial_- S)d\xi^{-}}Y^{i}= \sqrt{V}Y^{i}  \ \ , \ \ \partial_k V\partial_-\overline{X}^{k}
= \partial_-V
\ee

we get
\be
\partial_-w^i + \frac{1}{2}[(\partial_j S)w^j \partial_-\overline{X}^{i}- \partial_l S\delta^{il}\delta_{jk} \partial_- \overline{X}^k w^j] = H^{i}(\xi^-)
\ee 

Here $H^i = \sqrt{V}h^i$ . In terms of vector notation, this can be written as
\be
\frac{d\vec{w}}{d\xi^-} + \frac{1}{2}\vec{w}\times \vec{B} = H^i(\xi^-)
\ee

where $\vec{B}= \partial_-\vec{X} \times \nabla S$. We first consider a circular profile
\be
\overline{X}_1 = R \cos\xi^- \ \ \ \ \ , \ \  \  \ \ \overline{X}_2 = R \sin\xi^- 
\ee

Then we get following equations
\begin{eqnarray}
\frac{d w^{(1)}}{d\xi^-}  + \frac{Q}{2VR}w^{(2)} = H^1(\xi^-) \\
\frac{d w^{(2)}}{d\xi^-}  -\frac{Q}{2VR}w^{(1)} = H^1(\xi^-) 
\end{eqnarray}

Solution to these equations is 
\begin{eqnarray}
w^{(1)} + i w^{(2)} = C_1 (\xi^+) e^{-i\alpha \xi^-}  + G^{(1)}(\xi^-) \\
w^{(1)} -i w^{(2)} = C_2 (\xi^+) e^{i\alpha \xi^-} + G^{(2)}(\xi^-)
\end{eqnarray}

Here $\alpha= \frac{Q}{2VR}$ and $G_1 ,G_2$ are arbitrary functions. Now we consider the case when in base configuration has non-constant radius. 
\be
\overline{X}_1 = R(\xi^-) \cos\xi^- \ \ \ \ \ , \ \  \  \ \ \overline{X}_2 = R(\xi^-) \sin\xi^- 
\ee

Now putting this in 
 \be
\partial_-w^i + \frac{1}{2}[(\partial_j S)w^j \partial_-\overline{X}^{i}- \partial_l S)\delta^{il}\delta_{jk} \partial_- \overline{X}^k w^j] = H^{i}(\xi^-)
\ee 

we get same equations
\begin{eqnarray}
\frac{d w^{(1)}}{d\xi^-}  + \frac{Q}{2VR}w^{(2)} = H^1(\xi^-) \\
\frac{d w^{(2)}}{d\xi^-}  -\frac{Q}{2VR}w^{(1)} = H^1(\xi^-) 
\end{eqnarray}

Only change is that $R=R(\xi^-)$. Terms containing derivatives of $R$ cancel. We can change the independent variable  to $\tilde{\xi}= \tilde{\xi}(\xi^-)$ such that
\be
\frac{d}{d\xi^-} = \frac{d\tilde{\xi}}{d\xi^-}\frac{d}{d\tilde{\xi}} \ \ \ \ , \ \ \ \  \frac{2VR}{Q}\frac{d\tilde{\xi}}{d\xi^-}=1
\ee
Then the equation becomes like constant coefficient case. New variable $\tilde{\xi}$ is given by
\be
\tilde{\xi} = \frac{Q}{2}\int \frac{d\xi^-}{Q+R(\xi^-)}
\ee

Solution is 
\begin{eqnarray}
w^{(1)} + i w^{(2)} = C_1 (\xi^+) e^{-i\tilde{\xi}}  + G^{(1)}(\tilde{\xi})\\
w^{(1)} -i w^{(2)} = C_2 (\xi^+) e^{i\tilde{\xi}} + G^{(2)}(\tilde{\xi})
\end{eqnarray}

Again $G_1 , G_2$ are arbitrary functions. 
\subsection{D0-F1-KK picture}

Now we consider same profile in $D0$-$NS1$ duality frame. In polar coordinates, we have

\begin{equation}
R = \overline{R} + \epsilon r  \ \ \ ,  \ \ \ E =1 + \epsilon \dot{a}_y \ \ \ , \ \ \ B = \overline{B} -\epsilon a'_y
\end{equation}

We have put $X_3=0$ or $\theta=\frac{\pi}{2}$. Putting these in Lagrangian \ref{lag}  and as in previous section, expanding upto second order in 
$\epsilon$, we get
\begin{eqnarray}
L^{(2)}= \frac{-T_2}{2\overline{B}}\left(\dot{r}^2(V_{0}^{2}\overline{R}^2 + V_0 \overline{B}^2)+ 
\dot{a}_y^{2}(V_0 \overline{R}^2 + \frac{V_{0}^{2}\overline{R}^4}{\overline{B}^2}) \right. \nonumber \\
+ \left. \frac{2\dot{a}_y a'_y V_0 \overline{R}^2}{\overline{B}} + (4 \overline{R}
r V_0 + 2Q r)\dot{a}_y + 2\overline{B}V_0 \dot{r}r' \right)
\eea

From this we get following equations of motion
\begin{eqnarray}
\frac{V_0 ^2 \overline{R}^2 + V_0 \overline{B}^2}{\overline{B}}\ddot{r} + 2V_0\partial_t\partial_{\sigma_2}r - \frac{\dot{
a}_y }{\overline{B}}(2V_0 \overline{R} - Q) =0 \\
\frac{V_0 ^2 \overline{R}^4 + V_0 \overline{R}^2\overline{B}^2}{\overline{B}^3}\ddot{a}_y + 
\frac{2V_0 \overline{R}^2}{\overline{B}^2}\partial_t\partial_{\sigma_2}a_y + \frac{\dot{
r}}{\overline{B}}(2V_0 \overline{R} - Q) =0 
\end{eqnarray}

Simplifying, we get
\begin{eqnarray}
\frac{V_0 \overline{R}^2 + \overline{B}^2}{\overline{B}}\ddot{r} + 2\partial_t\partial_{\sigma_2}r - \frac{2\overline{R}\dot{
a}_y }{\overline{B}}(1 - \frac{Q}{2V_0 \overline{R}}) =0 \\
\frac{V_0 \overline{R}^2 + \overline{B}^2}{\overline{B}}\ddot{a}_y + 2\partial_t\partial_{\sigma_2}a_y +\frac{2\overline{B}\dot{
r}}{\overline{R}}(1 - \frac{Q}{2V_0 \overline{R}}) =0 
\end{eqnarray}

As in flat space case, we see that only time derivatives of the perturbations $r$ and $a_y$ occur. Hence any static
deformation is a solution. Solution to above equations can be written as 
\begin{eqnarray}
r= c_1(\xi^+) \cos(1-\alpha)\sigma + c_2(\xi^+) \sin (1-\alpha)\sigma \\
a_y = -\frac{\overline{B}}{\overline{R}}\left( c_1(\xi^+) \cos(1-\alpha)\sigma + c_2(\xi^+) \sin (1-\alpha)\sigma \right)
\end{eqnarray}

Here $\alpha= \frac{Q}{2V_0 \overline{R}}$. 

\section{Near ring limit of Bena-Kraus metric}

Till now we have considered, DBI description of supertubes in KK monopole background. In this description, string coupling $g_s$ is zero and backreaction
of supertube on geometry is not considered. Now we increase $g_s$ so that we have a gravitational description (supergravity). A metric for 
$3$-charge system $D1$-$D5$-$KK$ was given by Bena and Kraus in \cite{bena1} and generalized to include more dipole charges in \cite{bena2,elvang1}. 
Since this description, corresponds to supertube in KK monopole background, we consider perturbations in this system. 
The type IIB string frame solution is \cite{bena1}
\bea
ds_{10}^2 = \frac{1}{\sqrt{Z_1 Z_5}}\left[-(dt+k)^2 + (dy-k-s)^2\right] + \sqrt{Z_1Z_5}ds^2_K + \sqrt{\frac{Z_1}{Z_5}}ds^2_{T^4} \\
e^{\Phi}= \sqrt{\frac{Z_1}{Z_5}} \ \ \ \ \ \ , \ \ \ \ \ \ \ \  F^{(3)}= d[Z_1^{-1}(dt+k)\wedge (dy -s-k)] -*_4dZ_5
\eea

where $*_4$ is taken with respect to the metric $ds^2_{K}$ and 
\bea
ds^2_K = Z_K (dr^2 + r^2 d\theta^2 + r^2 \sin^2\theta d\phi^2) + \frac{1}{Z_K}(R_Kd\psi + Q \cos\theta d\phi)^2 \\
Z_K = 1+ \frac{Q}{r} \ \ , \ \ Z_{1,5} = 1+ \frac{Q_{1,5}}{\Sigma} \ \ , \ \ \Sigma = \sqrt{r^2 + R^2 + 2R r\cos\theta} 
\eea

From singularity analysis, bena-kraus derived following periodicity condition also
\be
y \cong y + 2\pi R_y \ \ , \ \  R_y = \frac{2\sqrt{Q_1 Q_5 \tilde{Z}_K}}{n} \ \ , \ \ \tilde{Z}_K = 1+ \frac{Q}{R}
\ee

One forms $s$ and $k$ have following components
\bea
s_{\psi} = - \frac{\sqrt{Q_1 Q_5 \tilde{Z}_K}R_K}{Z_K r \Sigma} \left[ \Sigma -r + \frac{r\Sigma}{Q\tilde{Z}_K} \right] \\
s_{\phi} = - \frac{\sqrt{Q_1 Q_5 \tilde{Z}_K}}{ \Sigma} \left[ R - \frac{(\Sigma -\frac{\Sigma}{Z_K} -r)}{Z_K}\cos\theta \right] \\
k_{\psi} =  \frac{\sqrt{Q_1 Q_5 \tilde{Z}_K}R_K Q}{2R \tilde{Z}_K Z_K r \Sigma} \left[ \Sigma -r -R -\frac{2rR}{Q} \right] \\
k_{\phi} = - \frac{\sqrt{Q_1 Q_5 \tilde{Z}_K}Q}{2R \tilde{Z}_K \Sigma} \left[ \Sigma -r -R + \frac{\Sigma -r +R}{\tilde{Z}_K}\cos\theta \right] \\
\eea

Chrages are quantized according to 
\be
Q = \frac{1}{2}N_K R_K \ \ , \ \ Q_1 = \frac{(2\pi)^4 g \alpha'^{3}N_1}{2R_K V_4} \ \ , \ \ Q_5 = \frac{g\alpha' N_5}{2R_K}
\ee

These coordinates are centred at KK monopole. It would be much more arduous task to consider perturbations in full geometry above. So as in 
\cite{Giusto}, we take near ring or thin tube limit of the above geometry. To take the near ring limit, it's better to use coordinates 
centred on ring. We define new coordinates by
\be
\rho = \Sigma \ \ , \ \  \phi=\phi \ \ , \ \ \psi=\psi \ \ , \ \ \cos\theta_1 = \frac{R + r\cos\theta}{\Sigma}
\ee

Another way to write is 
\be
\rho \sin \theta_1 = r \sin\theta \ \ , \ \ r^2 = \rho^2 + R^2 -2R\rho\cos\theta_1 
\ee

It's easy to see that for this change of coordinates
\be
dr^2 + r^2 d\theta^2 + r^2 \sin^2\theta = d\rho^2 + \rho^2 d\theta_1 ^2 + \rho^2 \sin^2 \theta_1 d\phi^2
\ee

We want to take the limit $R \rightarrow \infty$ keeping $\rho,\theta_1$ fixed. It's easy to see that under this limit
\be
r \sim R \ \ , \ \ \cos\theta \rightarrow -1 \ \ , \ \ Z_K \rightarrow 1  \ \ , \ \ \tilde{Z}_K \rightarrow 1
\ee

Now $Z_{1,5} = 1+ \frac{Q_{1,5}}{\rho}$ and $ds^2_K$ becomes
\be
(R_K d\psi -Q d\phi)^2 + d\rho^2 + \rho^2 d\theta_1 ^2 + \rho^2 \sin^2 \theta_1 d\phi^2
\ee

Defining $z= R_K \psi -Q \phi$, we see that this is a metric for $R^{3}\times S^{1}$. 
One forms becomes
\bea
k_\psi = -\frac{\sqrt{Q_1Q_5}R_K}{\rho} \\
k_\phi = \frac{\sqrt{Q_1Q_5}Q }{\rho} \\
s_{\psi}= \frac{\sqrt{Q_1Q_5}R_K}{\rho} - \frac{\sqrt{Q_1Q_5}R_K}{Q} \\
s_{\phi} = -\frac{\sqrt{Q_1Q_5}Q}{\rho} - \sqrt{Q_1Q_5}\cos\theta_1
\eea

Now combining them, we get
\bea
k = k_\psi d\psi + k_\phi d\phi = -\frac{\sqrt{Q_1Q_5}}{\rho}(R_K d\psi -Q d\phi) \\
k+ s = (k_\psi + s_\psi)d\psi + (k_\phi + s_\phi)d\phi = -\frac{\sqrt{Q_1Q_5}R_K}{Q}d\psi -\sqrt{Q_1Q_5}\cos\theta_1 d\phi
\eea

Defining $z= R_K \psi -Q\phi$, we have
\be
k = -\frac{\sqrt{Q_1Q_5}}{\rho}dz \ \ \ , \ \ \ -(k+s) = \frac{\sqrt{Q_1 Q_5}}{Q}dz  + \sqrt{Q_1Q_5}(1-\cos\theta)d\phi
\ee

In terms of these quantities, the metric becomes
\dm
ds_{10}^2 = \frac{1}{\sqrt{Z_1 Z_5}}\left[-(dt -\frac{\sqrt{Q_1Q_5}}{\rho}dz )^2 + (dy +\frac{\sqrt{Q_1 Q_5}}{Q}dz  + 
\sqrt{Q_1Q_5}(1-\cos\theta_1)d\phi)^2\right] 
\edm
\be
+ \sqrt{Z_1Z_5}ds^2_K + \sqrt{\frac{Z_1}{Z_5}}ds^2_{T^4}
\ee

Define $\tilde{y}= y +\frac{\sqrt{Q_1 Q_5}}{Q}z$ and then we see that it is same as near ring limit of Maldacena-Maoz
\dm
ds_{10}^2 = \frac{1}{\sqrt{Z_1 Z_5}}\left[-(dt -\frac{\sqrt{Q_1Q_5}}{\rho}dz )^2 + (d\tilde{y}  + 
\sqrt{Q_1Q_5}(1-\cos\theta_1)d\phi)^2\right] 
\edm
\be
+ \sqrt{Z_1Z_5}ds^2_K + \sqrt{Z_1}{Z_5}ds^2_{T^4}
\ee

KK monopole structure fixes periodicity of $\tilde{y}$. 
\be
R_{y}= \frac{2\sqrt{Q_1Q_5}}{n}
\ee

Now we write down the RR field. We have
\be
F^{(3)}= d[Z_1^{-1}(dt+k)\wedge (dy -s-k)] -*_4dZ_5
\ee

Since
\be
dZ_5 = -\frac{Q_5}{\rho^2} d \rho  
\ee

we have
\be
*_4 dZ_5 = -\frac{Q_5}{\rho^2} \sqrt{g_4}g^{\rho\rho}\epsilon_{\rho\theta_1\phi z}d\theta_1 \wedge d\phi \wedge dz = -Q_5 \sin
\theta_1 d\theta_1 \wedge d\phi \wedge dz
\ee

where we have put $\epsilon_{\rho\theta_1\phi z}=1$ as in KK monopole space. Since four dimensional base space is $R^3 \times
S^1$ we have used flat metric. 
Writing 
\be
\sigma = dy + \sqrt{Q_1 Q_5}(1-\cos\theta_1) \ \ \ , \ \ \ d\sigma = \sqrt{Q_1Q_5}\sin\theta_1 d\theta_1 \wedge d\phi
\ee

we can write 
\be
*_4 dZ_5 = -d(Q_5 \sigma \wedge dz)
\ee

So $F^{(3)}=dC^{(2)}$ where
\be
C^{(2)}= \frac{1}{Z_1}\left[(dt-\frac{\sqrt{Q_1Q_5}}{\rho})\wedge \sigma \right] +Q_5 \sigma \wedge dz
\ee

We see that only effect of KK monopole on supertube geometry is that of compactifying $R^4$ to $R^3 \times S^1$ with 
radius of $S^1$ determined in terms of KK monopole charges. One can easily dualize this to 'thin tube' limit of $NS1$-$P$ system. So the results of \cite{Giusto} involving near ring
limit can be taken over for this system. In \cite{Giusto}, we only considered fluctuations along torus directions when considering
 near ring or 'thin tube' limit of $NS1$-$P$ geometry. In our case here, period of oscillation on gravity side will be same as period calculated in \cite{Giusto}.
 In terms of masses, gravity calculation gave
 \be
 \Delta t = \frac{1}{2T_{NS1}} (M_{NS1} + M_P )
 \ee
This is same as equation \ref{period} from DBI analysis after one does the dualities. For fluctuations in torus directions in DBI limit, there is no difference between flat background and KK monopole background except that
 $R_{0}^2$ for supertube would be calculated using KK monopole metric. So period of oscillations match. Fluctuations along taub-nut directions 
are difficult to solve and we postpone that work to a future publication.

\section{ Results and discussion}

We studied supertubes in various profiles moving in a KK monopole background. At the DBI level, profile which corresponds to Bena-Kraus metric in gravity limit
was analyzed in both $NS1$-$P$ and $D0$-$NS1$ duality frames. We considered perturbations of supertube with this profile and found that motion of 
supertube in KK monopole background is not a drifting motion but more like quasi-oscillations as considered in \cite{Giusto}. This can be taken as
 evidence for the bound state nature of system corresponding to BK profile. But since conjecture of \cite{Giusto} was based on flat background geometry
 (at DBI level), one should be cautious in considering this as definitive for the bound state nature of the system.

Near ring or thin tube limit of $D1$-$D5$-$KK$
 turned out to be identical to near ring limit of $D1$-$D5$ supertube alone, only change occuring in the periodicity of the ring circle. 
In our present case
 of $D1$-$D5$-$KK$, the periodicity of the ring is determined by the monopole charge while with just $D1$-$D5$, it could be arbitrary. 
Calculations of period of oscillation at DBI level, for torus directions, match with the gravity analysis. Both 
are very similar to $D1$-$D5$ supertube case dealt in \cite{Giusto}. Only difference comes from the fact that in  $D1$-$D5$-$KK$ case, radius of 
supertube is calculated using KK monopole metric rather than flat metric.

Substantial difference from flat space case occurs when one considers form of fluctuation
 and not just the periodicity. In KK monopole background, even at linear level, there is no separation of dependences on $\xi^+$ and $\xi^-$ and thus
 left-movers and right-movers are invariably mixed. This effect is due to curvature of background. We also analyzed, at DBI level, perturbations which
  have profile functions different from BK profile even though in these cases no gravity description is known and so can not be compared with DBI analysis. 
 
It would be interesting to analyze $KK$-$P$ system constructed in \cite{self} using the linearized perturbation formalism as developed in the present paper.
Since we know that $KK$-$P$ is a bound state, it can give us insight about how the conjecture of \cite{Giusto} works in presence of KK monopoles. One can also
work with other three charge systems which do not contain KK monopole like $D1$-$D5$-$P$ system constructed in \cite{gms1,gms2} and are known to be bound states.

\section{ Acknowledgements}

I would like to thank my advisor Samir Mathur for several helpful discussions and support at various stages of the project.
I would also like to thank Stefano Giusto and Ashish Saxena for several discussions and help with the manuscript.

\end{document}